\documentclass[]{elsart}
\usepackage{dcolumn}% Align table columns on decimal point
\usepackage{bm}% bold math
\usepackage{indentfirst}
\usepackage{color}
\usepackage{graphicx}
\usepackage{amsmath}
\usepackage{amssymb}

\newcommand{\g}{\gamma}
\newcommand{\nslash}{\kern 0.2 em n\kern -0.50em /}
\newcommand{\kslash}{\kern 0.2 em k\kern -0.45em /}
\newcommand{\pslash}{\kern 0.2 em p\kern -0.50em /}
\newcommand{\Sslash}{\kern 0.2 em S\kern -0.50em /}
\newcommand{\Pslash}{\kern 0.2 em P\kern -0.50em /}
\newcommand{\ii}{i}
\newcommand{\eps}{\epsilon}
\newcommand{\slim}{\mskip 1.5mu}              % small space in math
\newcommand{\Tr}{\operatorname*{Tr}\nolimits} % Trace operator

\newcommand{\cdott}{{\mskip -1.5mu} \cdot {\mskip -1.5mu}}

\newcommand{\bi}{\begin{itemize}}
\newcommand{\ei}{\end{itemize}}
\newcommand{\be}{\begin{equation}}
\newcommand{\ee}{\end{equation}}
\newcommand{\ba}{\begin{eqnarray}}
\newcommand{\ea}{\end{eqnarray}}

\journal{Physics Letters B}

\begin{document}

\begin{frontmatter}

\title{T-odd quark-gluon-quark correlation function in the diquark model}

\author[seu]{Zhun Lu\corauthref{cor}}
\corauth[cor]{Corresponding author at: Department of Physics, Southeast University, Nanjing
211189, China.} \ead{zhunlu@seu.edu.cn},
\author[utfsm]{Ivan Schmidt}
\address[seu]{Department of Physics, Southeast University, Nanjing
211189, China}
\address[utfsm]{Departamento de F\'\i sica, Universidad T\'ecnica
Federico Santa Mar\'\i a, and Centro Cient\'\i fico-Tecnol\'ogico de
Valpara\'\i so Casilla 110-V, Valpara\'\i so, Chile}

\begin{abstract}
We study the transverse momentum dependent quark-gluon-quark correlation function.
Using a spectator diquark model, we calculate the eight time-reversal-odd interaction-dependent twist-3 quark distributions
appearing in the decomposition of the transverse momentum dependent quark-gluon-quark correlator.
In order to obtain finite results, we assume a dipole form factor for the nucleon-quark-diquark
coupling, instead of a point-like coupling.
The results are compared with the time-reversal-odd interaction-independent twist-3 TMDs
calculated in the same model.
\end{abstract}

\end{frontmatter}

\section{Introduction}
In the theoretical description of high energy (semi-)inclusive processes involving hadrons,
the cross-sections are usually expanded in powers of $1/Q$, where $Q$ is the large
momentum transfer of the collision.
The contribution at the leading power can be expressed as a convolution of the
leading-twist (or twist-two) distribution/fragmentation functions and the hard
scattering coefficients.
In the first subleading power of the $1/Q$ expansion, the twist-3 distribution
and/or fragmentation functions contribute to the cross-section~\cite{twist3}.
Unlike the twist-two distribution functions that describe the parton densities
inside the nucleon, there is no probability interpretation for the
twist-3 distribution functions.
However, they provide a wealth of information about the nucleon parton structure~\cite{Bacchetta:2006tn},
especially when the parton transverse momenta are present.
The interest on the twist-3 contributions also comes from the fact that they
are related to the multi-parton correlation inside the nucleon~\cite{Jaffe:1989xx,
Burkardt:2008ps}.

In this letter we apply a phenomenological model to study twist-3 quark distributions
that are encoded in the quark-gluon-quark correlation.
We pay special attention on the time-reversal-odd (T-odd) transverse momentum
dependent (TMD) distributions.
In the TMD factorization approach, the leading twist T-odd TMD~\cite{sivers}
play important roles~\cite{bhs02,collins02,jy02}
in the single-spin asymmetries (SSAs) measured in semi-inclusive deeply inelastic
scattering (SIDIS)~\cite{hermes10,Alekseev:2010rw, Qian:2011py}.
At the twist-3 level, much richer phenomena arise, i.e. there are eight T-odd distributions
that can contribute to various azimuthal asymmetries in the SIDIS~\cite{Bacchetta:2006tn} and Drell-Yan~\cite{Lu:2011th} processes.
Although the twist-3 contributions are suppressed by $1/Q$, they are potential
experimental observables and may be accessible in the kinematical regime where $Q$
is not so large.
The experiments at Jefferson Lab~\cite{Avakian:2003pk,Aghasyan:2011ha} and PAX~\cite{PAX_05}
are ideal for exploring this kinematical region.

\section{Interaction-dependent T-odd twist-3 distributions}
Our starting point is the transverse momentum dependent quark-gluon-quark
correlation function, which is defined as~\cite{bmp03}
\begin{equation}
\begin{split}
& \left( \tilde\Phi^{[\pm]\alpha}_A\right)_{ij}
(x,p_T)
\equiv
\int \frac{d^{2}{\xi_T}d{\xi^-}}{(2\pi)^3}
e^{ip\xi}
\\
&\phantom{=}\ \times
\langle P,S| \bar{\psi}_j(0)
g\int_{\pm \infty}^{\xi^-}
 d\eta^- \mathcal{L}^{[\pm]}(0,\eta^-)
F^{+\alpha}(\eta) \mathcal{L}^{\xi_T\!,\ \xi^+}(\eta^-,\xi^-)
\psi_i(\xi) |P,S \rangle_\text{c}
 \bigg|_{\begin{subarray}{l}
\eta^+ = \xi^+=0 \\ \eta_T = \xi_T \\ p^+ = xP^+ \end{subarray}}\,,
\end{split}
\label{qqg}
\end{equation}
where $F^{\mu\nu}$ is the antisymmetric field strength tensor of the gluon,
$\mathcal{L}^{[\pm]}$ and $\mathcal{L}^{\xi_T\!,\ \xi^+}$ are the gauge-links
ensuring the gauge-invariance of the definition.
The sign ``$\pm$" in the superscript or subscript indicates that the
gauge-link between the quark and the gluon is future/past-pointing~\cite{collins02},
corresponding to the SIDIS/Drell-Yan processes, respectively.

The correlator in Eq.~(\ref{qqg}) can be rewritten further to
\begin{equation}
\begin{split}
&\left( \tilde\Phi^{[\pm]\alpha}_A\right)_{ij}
(x,p_T) =
ig\int \frac{d^{2}{\xi_T}d{\xi^-}d\eta^-}{(2\pi)^4}
 \int dx^\prime  {e^{i x^\prime P^+ \eta^- }\over (x^\prime \mp i\epsilon)}
e^{i[(x-x^\prime)P^+ \cdot \xi^-- \bm p_T \cdot \bm{\xi_T}]}
\\
& \quad\times\langle P,S| \bar{\psi}_j(0)
\mathcal{L}^{[\pm]}(0,\eta^-) F^{+\alpha}(\eta) \mathcal{L}^{\xi_T\!,\ \xi^+}
(\eta^-,\xi^-) \psi_i(\xi) |P,S \rangle
 \bigg|_{\begin{subarray}{l}
\eta^+ = \xi^+=0 \\ \eta_T = \xi_T \end{subarray}}\,.
\end{split}
\label{qqgsimp}
\end{equation}

If the parton transverse momentum is integrated over, one can define the
collinear quark-gluon correlator, the so-called Efremov-Teryaev-Qiu-Sterman
(ETQS) function~\cite{twist3,Kouvaris:2006zy}, as
\begin{eqnarray}
T_{q,F}(x,x) & =&
\int \frac{d\xi^-d\eta^-}{4\pi}
e^{ixP^+ \xi^-}\nonumber\\
&\times&
\langle P,S| \bar{\psi}(0)
  \mathcal{L}(0,\eta^-)
\gamma^+[\epsilon^{\alpha S_T}F_{\alpha}^{\,\,+}(\eta^-)]
\mathcal{L}(\eta^-,\xi^-) \psi_i(\xi^-) |P,S \rangle\,,
\label{etqs}
\end{eqnarray}
where $\epsilon^{\alpha S_T}=\epsilon^{\alpha\rho} S_{T\rho}$, with $S_T$ the
transverse polarization vector of the nucleon.
In the twist-3 collinear factorization approach~\cite{twist3}, a nonzero ETQS
function has been applied to explain the large single transverse spin
asymmetries (SSAs) observed in $p^\uparrow p \rightarrow \pi X$
processes~\cite{ssa}.
According to Eq.~(\ref{qqgsimp}), the ETQS function can be obtained from the
transverse momentum dependent correlator $\tilde\Phi^{[+]}_{A\, \alpha}$ by
\begin{equation}
gT_{q,F}(x,x)  =\epsilon^{\alpha S_T}\int d^2 p_T\text{Tr}\left( \gamma^+
\tilde\Phi^{[+]}_{A\, \alpha}\big{|}_{\text{G.P.}}(x,p_T)\right)\,,
\end{equation}
where $\tilde\Phi^{[+]}_{A\, \alpha}\big{|}_{\text{G.P.}}$ denotes the gluonic
pole part of $\tilde\Phi^{[+]}_{A\, \alpha}$, which can be obtained by taking
the imaginary component of the factor $1/(x^\prime\mp i\epsilon)$
in Eq.~(\ref{qqgsimp}):
\begin{equation}
{1\over (x^\prime \mp i\epsilon)} =
\text{P}\left({1\over x^\prime}\right)\pm i \delta (x^\prime)\,.\label{fac}
\end{equation}

The quark-gluon-quark correlator can be decomposed as~\cite{Bacchetta:2006tn}
\begin{align}
\tilde\Phi_A^{\alpha}(x,p_T) =
 & \frac{x M}{2}\,
\biggl\{
\Bigl[
\bigl(\tilde{f}^\perp-\ii\slim \tilde{g}^{\perp} \bigr)
        \frac{p_{T \rho}^{}}{M}
-\bigl(\tilde{f}_T'+ \ii\slim \tilde{g}_T'\bigr)
     \,\epsilon_{T \rho\sigma}^{}\slim S_{T}^{\sigma}
\nonumber \\ &
-\bigl(\tilde{f}_s^{\perp}+\ii\,\tilde{g}_s^{\perp}\bigr)
     \frac{\epsilon_{T \rho \sigma}^{}\slim p_{T}^{\sigma}}{M}
 \slim\Bigl]
\bigl(g_T^{\alpha \rho} - i \epsilon_T^{\alpha\rho} \gamma_5\bigr)
-\bigl(\tilde{h}_s + \ii\,\tilde{e}_s\bigr)
        \gamma_T^{\alpha}\,\gamma_5\nonumber \\
 &
%\nonumber \\[0.2em] & \qquad
+\Bigl[\bigl(\tilde{h} + \ii\,\tilde{e}\bigr)
  +\bigl( \tilde{h}_T^{\perp}
        - \ii\,\tilde{e}_T^{\perp}\bigr)\slim
   \frac{\eps_T^{\rho \sigma} p_{T\rho}^{}\slim S_{T\sigma}^{}}{M}
 \slim\Bigr]
  \ii \gamma_T^{\alpha}
\biggr\} \frac{\nslash_+}{2}\,.
\label{eq:phiAalpha}
\end{align}
The functions appearing with a tilde in the above expression are the {\it interaction-dependent} twist-3 quark distributions, which depend also on the longitudinal momentum fraction $x$ and the transverse momentum $p_T$.
Among them, $\tilde{e}$, $\tilde{f}^\perp$,
$\tilde{g}_T$ (or $\tilde{g}_T^\prime$), $\tilde{g}_T^{\perp}$, $\tilde{g}_L^\perp$,
$\tilde{h}_L$, $\tilde{h}_T$ and $\tilde{h}_T^{\perp}$ are T-even; and  $\tilde{e}_L$, $\tilde{e}_T$, $\tilde{e}_T^\perp$,
$\tilde{f}_T$ (or $\tilde{f}_T^\prime$), $\tilde{f}_T^\perp$, $\tilde{f}_L^\perp$,
$\tilde{g}^\perp$ and $\tilde{h}$ are T-odd.
All the TMDs can be projected out by different Dirac matrices:
\begin{align}
\frac{1}{2M x} \Tr \big[\tilde{\Phi}_{A\slim \alpha}\,
  \sigma^{\alpha+} \big]
&= \tilde{h}  + i\,\tilde{e}
  + \frac{\eps_T^{\rho \sigma} p_{T\rho} S_{T\sigma}}{M}\,
    \bigl(\tilde{h}_T^{\perp}
  - i\,\tilde{e}_T^{\perp}\bigr),\label{trace1}
\\
\frac{1}{2M x} \Tr
\big[ \tilde{\Phi}_{A\slim \alpha}\,
  \ii \sigma^{\alpha+} \gamma_5 \big]
&=  S_L\,\bigl(\tilde{h}_L+ i\,\tilde{e}_L\bigr)
  - \frac{p_{T} \cdott S_{T}}{M}\,
  \bigl(\tilde{h}_T + i\,\tilde{e}_T\bigr),\label{trace2}
\\
\frac{1}{2M x}\,
  \Tr \big[\tilde{\Phi}_{A \rho}\slim (g_T^{\alpha\rho}
          + i \epsilon_T^{\alpha\rho} \gamma_5)\slim \gamma^+ \big] &=
\frac{p_T^\alpha}{M}
        \bigl(\tilde{f}^\perp - i \tilde{g}^\perp\bigr)
- \epsilon_T^{\alpha\rho}\slim S_{T\rho}^{}\,
        \bigl(\tilde{f}_T + i \tilde{g}_T\bigr)
\nonumber \\ &\quad
\hspace{-5cm}
- S_L \,\frac{\epsilon_T^{\alpha\rho}\slim p_{T\rho}^{}}{M} \,
        \bigl(\tilde{f}_L^\perp + i\,\tilde{g}_L^\perp\bigr)
%\nonumber \\ &\quad
-      \frac{p_T^{\alpha}\,p_T^{\rho}
     -\frac{1}{2}\,{p}_T^{2}\,g_{T}^{\alpha\rho}}{M^2}\,
        \eps_{T \rho\sigma}^{}\slim S_{T}^{\sigma}
        \,  \bigl(\tilde{f}_T^{\perp} + i \tilde{g}_T^{\perp}\bigr)\,.\label{trace3}
\end{align}

Taking the real part of the right hand side of Eq.~(\ref{fac}),
one can deduce from the traces the T-even TMDs;
if one takes the imaginary part of the right hand side of Eq.~(\ref{fac}) instead,
one can obtain the T-odd TMDs.
In this work we will study the possibility to calculate these T-odd TMDs from specific models.
In a realistic calculation we ignore all the gauge-links in the correlator
(\ref{qqg}),  as a lowest order approximation.
Also, we choose the diquark model~\cite{Jakob:1997wg,Bacchetta:2008af}, which
has been widely applied in the calculation of TMD
distributions, and it is the first model that shows that the T-odd distributions
are non-vanishing.
For simplicity we only consider the case that the diquark is a scalar,
since the extension to the spectator model with (axial-)vector diquarks is
straight forward.
In the following we will calculate the T-odd TMDs appearing in the DIS process,
since those in Drell-Yan process are related to them by a minus sign, as indicated
by Eq.~(\ref{fac}).

\begin{figure}
\begin{center}
\includegraphics[scale=0.55]{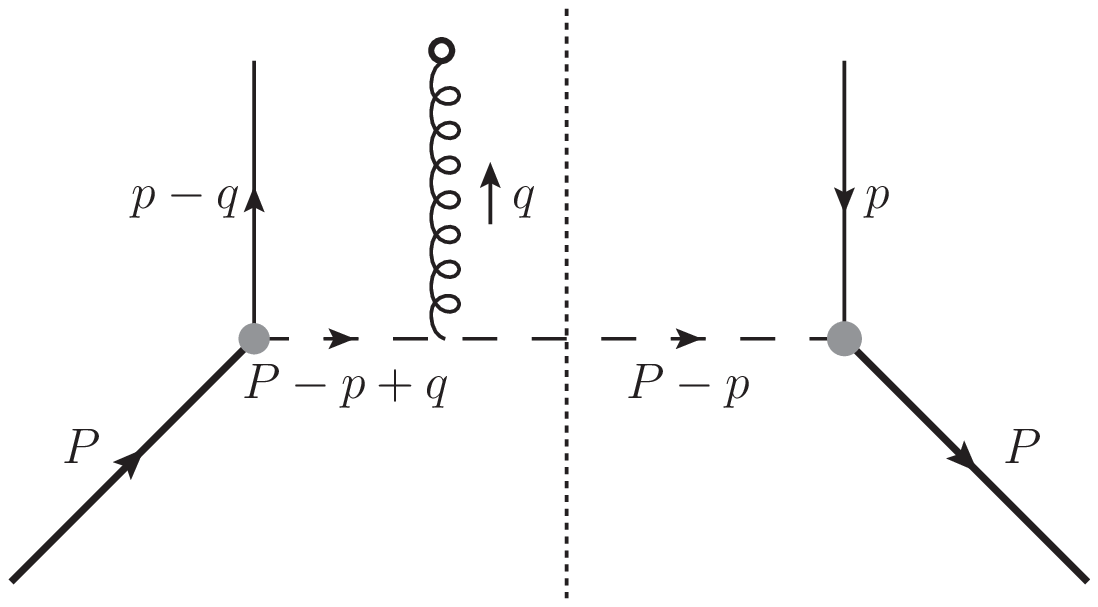}
\includegraphics[scale=0.55]{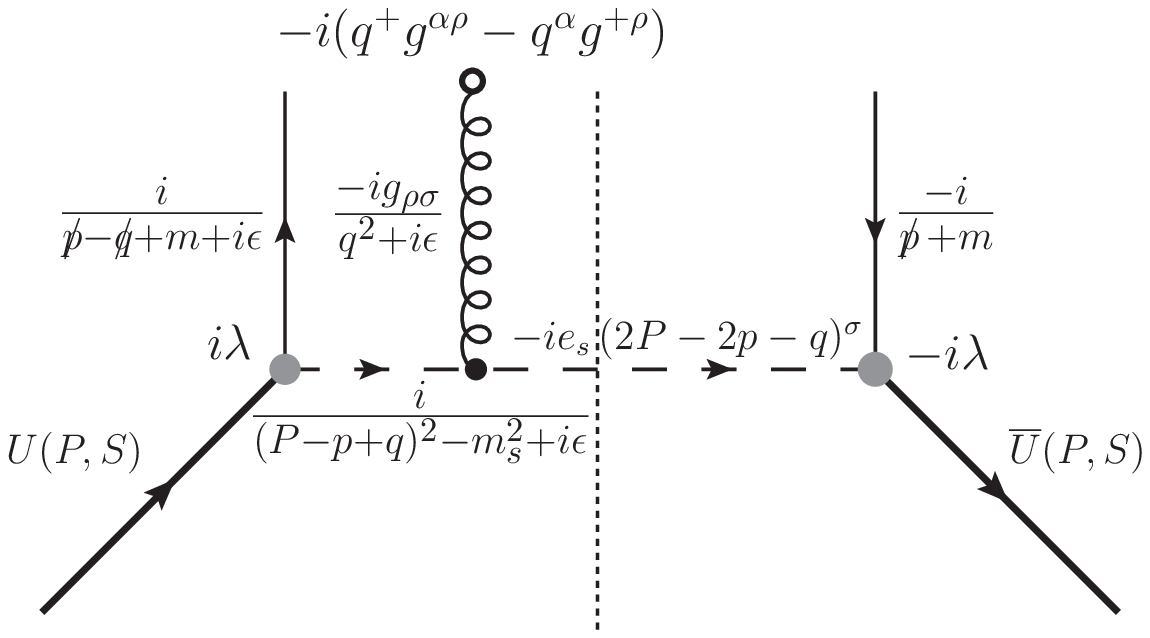}
\caption{
Diagram to calculate the quark-gluon-quark correlator in the diquark model.
The momenta of the nucleon, the quark and the diquark are denoted by $P$, $p$ and $q$,
respectively.
In the right panel we show the Feynman rules applied in the calculation.
The open circle on the upper end of the gluon lines indicates a special Feynman rule for
the field strength tensor in the definition of quark-gluon-quark correlator, see also Refs.~\cite{Collins:1981uw,Goeke:2006ef}} \label{spec}
\end{center}
\end{figure}

The Lagrangian for the scalar diquark model with Abelian gauge field (electromagnetism) reads:
\begin{eqnarray}
 \mathcal{L}(x)
 &=&\bar\Psi(x)\,(i\gamma^\mu\partial_\mu-M)\,\Psi(x)
  +\bar\psi(x)\,(i\gamma^\mu D_{q,\mu}-m_q)\,\psi(x)\nonumber\\
  &&+\varphi^*(x)\,( \overleftarrow{D}_s^{\mu*}\,\overrightarrow{D}_{s,\mu}-m_s^2)\,\varphi(x)\notag
 -\tfrac{1}{4}F^{\mu\nu}(x)\,F_{\mu\nu}(x)\nonumber\\
 &\quad&+\lambda\,\big[\bar\psi(x)\,\Psi(x)\,\varphi^*(x)+\bar\Psi(x)\,\psi(x)\,\varphi(x)\big] \,,\label{sdm}
\end{eqnarray}
where $\lambda$ denotes the nucleon-quark-diquark coupling, and
\begin{eqnarray}
 D_q^\mu\,\psi(x)=\big[\partial^\mu+ie_q\,A^\mu(x)\big]\,\psi(x), \quad
 D_s^\mu\,\varphi(x)=\big[\partial^\mu+ie_s\,A^\mu(x)\big]\,\varphi(x)\,,
\end{eqnarray}
here $e_q$ and $e_s$ are the charges for the quark and the spectator diquark, and
$M$, $m$ and $m_s$ denote the masses for the nucleon, the quark and the scalar diquark,
respectively.
Our calculation was performed in the Feynman gauge.
We also checked the calculation in the light-cone gauge (corresponding to
the gluon polarization sum $d^{\mu\nu}=-g^{\mu\nu}+(q^\mu n_-^\nu+ q^\nu n_-^\mu)/q^+$ ),
and found that it leads to the same results as in the Feynman gauge.
The Feynman diagram for calculating the correlator defined in the scalar diquark is shown
in Fig.~\ref{spec}, in the right panel of which we list explicitly the Feynman rules of the
vertices, propagators and external lines derived from the Lagrangian in (\ref{sdm}).
The vertical dashed line cutting the external diquark imposes the onshell condition
$\delta((P-q)^2-m_s^2)$.
When the calculation for the quark-gluon-quark correlator is transformed from the real QCD
to an Abelian theory, one should use the replacement $g\rightarrow - e_q$.
Also a specific Feynman rule~\cite{Collins:1981uw,Goeke:2006ef} (depicted by the open circle
at the end of the gluon line in Fig.~\ref{spec}) for the field strength tensor $F^{+\alpha}$
has been applied: $-i(q^+ g^{\alpha\rho}-q^\alpha g^{+\rho})$.

In the simplest version of the diquark model~\cite{bhs02}, the coupling $\lambda$ is
treated as a constant.
However, we find that in this simplest model there are divergences emerging for some
T-odd TMDs when one performs the integrations over the transverse momentum.
Similar divergences have already appeared explicitly in the calculation of
interaction-independent twist-3 T-odd TMDs in the same model~\cite{Gamberg:2006ru}.
We conclude that this is a common feature for the T-odd twist-3 distributions
when one applies the point-like coupling for the nucleon-quark-diquark interaction
vertex.
In order to obtain finite results, instead of a point-like coupling constant
$\lambda$, we choose a dipole form factor for the nucleon-quark-diquark
coupling~\cite{Jakob:1997wg}:
\begin{align}
\lambda\rightarrow \lambda(p^2) = {N_s(p^2-m^2) \over (p^2 -\Lambda^2)^2 },\label{dipole}
\end{align}
where $N_s$ is the normalization constant, $m$ is the mass for the quark,
and $\Lambda$ is the cut-off parameter for the quark momentum.
The same choice has been applied to calculate the leading-twist T-odd TMDs
for the nucleon~\cite{Bacchetta:2003rz} and the pion~\cite{Lu:2005rq}.

Applying the scalar diquark model in Eq.~(\ref{sdm}) to the quark-gluon-quark
correlator (\ref{qqgsimp}) and using the traces in Eqs.~(\ref{trace1},
\ref{trace2}, \ref{trace3}),
we obtain non-zero results for the eight T-odd interaction dependent
twist-3 quark distributions.
Here we refrain from given further details on our calculation, and only write down
the final expressions for these TMDs:
\begin{align}
\tilde{e}_T^\perp  (x,p_T^{\,2})& = -{A\over 2 x} {L-(xM+m)^2 \over  L (L+\vec p_T^{\,2})^3}
+\frac{A}{2 x}{1\over \vec p_T^{\,2}(L+\vec p_T^{\,2})^2} \ln{\vec p_T^{\,2} +L\over L} ,
\displaybreak[0] \label{etperp}\\
\tilde{e}_L (x,p_T^{\,2})&= {A\over 2x}{(xM+m)\over M } {L-\vec p_T^{\,2}\over L (L+\vec p_T^{\,2})^3},
\displaybreak[0] \\
\tilde{e}_T (x,p_T^{\,2})&= \frac{A}{2 x}   {(m+xM)^2 +2L+\vec p_T^{\,2} \over L (L+\vec p_T^{\,2})^3}
-\frac{A}{4 x}{1\over \vec p_T^{\,2}(L+\vec p_T^{\,2})^2} \ln{\vec p_T^{\,2} +L\over L},
\displaybreak[0] \\
\tilde{h} (x,p_T^{\,2})&=-{A\over 2Mx}  {xM +m\over  L(L+\vec p_T^{\,2})^2},
\displaybreak[0] \\
\tilde{g}^\perp (x,p_T^{\,2})&= {A\over 2x}  {(m+xM)^2+\vec p_T^{\,2}\over L (L+\vec p_T^{\,2})^3},
\displaybreak[0] \\
\tilde{f}_L^\perp (x,p_T^{\,2})&=-\frac{A}{2 x}  {(xM+m)^2-\vec p_T^{\,2}\over L (L+\vec p_T^{\,2})^3},
\displaybreak[0] \\
\tilde{f}_T^\prime(x,p_T^{\,2})&= - {A\over 2 xM} {xM+m\over L (L+\vec p_T^{\,2})^2 },
\displaybreak[0] \\
\tilde{f}_T^\perp(x,p_T^{\,2})&= {A\over  x}  {M(xM+m)\over L (L +\vec p_T^{\,2})^3},
\displaybreak[0] \\
\tilde{f}_T (x,p_T^{\,2})&= - {A\over 2 xM} {xM+m\over (L +\vec p_T^{\,2})^3}\label{ft} ,
\end{align}
where $L=(1-x)\Lambda^2+xm_s^2-x(1-x)M^2$, and ${A}= (1-x)^3e_qe_s  N_s^2/(4(2\pi)^4) $.

In the following we point out several comments about our results.

(i) Our calculation shows that one can obtain nonzero results for the T-odd
interaction-dependent TMDs, even without the presence of the gauge-links connecting
the quark and gluon fields.

(ii) Among the eight T-odd TMDs, $\tilde{g}^\perp$ and $\tilde{f}_L^\perp$ are
finite when one chooses a point-like nucleon-quark-diquark coupling\footnote{we find
that in the integrands of calculating $\tilde{g}^\perp$ and $\tilde{f}_L^\perp$,
the power of $q_T$ appearing is one, after integrating over $q^+$ and $q^-$.
Therefore, they are finite in the case of pointlike coupling.},
which are different to their interaction dependent partners $g^\perp$ and $f_L^\perp$.
The other six TMDsare divergent in this case. For consistency, all the TMDs are
calculated by adopting a dipole form factor for
the nucleon-quark-diquark coupling. The need of form factor to regularize the light-cone divergence in
certain twist-3 T-odd correlator has also been pointed out in
Ref.~\cite{Gamberg:2008yt}.

(iii) The results we present here are at the order $\mathcal{O}(e_qe_s)$.
We recall that in the diquark model the calculation of leading-twist T-odd TMDs
(i.e., the Sivers function and the Boer-Mulders function~\cite{Boer:1997nt})
in the one-gluon exchange approximation, the results are at the same order.
Therefore, our calculation can be compared with the one for
the Sivers function.
To do this, we evaluate the ETQS function from the
transverse momentum dependent quark-gluon-quark correlator (\ref{qqg}) in the
scalar diquark model:
\begin{eqnarray}
e_q T_{q,F}(x,x) =-\vec S_T^{\, 2}{\pi A(xM+m)\over 2 L^2}\,.
\end{eqnarray}
The first transverse-moment of the Sivers function in SIDIS in the scalar diquark model is~\cite{Bacchetta:2003rz}:
\begin{eqnarray}
f_{1T}^{\perp(1)}(x) =\int d^2 p_T \,{\vec p_T^{\, 2}\over 2M^2}\,f_{1T}^\perp (x,\vec p_T^{\, 2})
= {\pi A(xM+m)\over 4 M L^2}\,.
\end{eqnarray}
Therefore, in the diquark model at the order $\mathcal{O}(e_q e_s)$, we verify the famous identity~\cite{bmp03,mw04,tmdtw3,Kang:2011hk} between $T_{q,F}(x,x)$ and $f_{1T}^{\perp(1)}(x)$ (assuming
$\vec S_T^{\, 2}=1$):
\begin{eqnarray}
e_q T_{q,F}(x,x) = - 2M f_{1T}^{\perp(1)}(x)\,\label{relation},
\end{eqnarray}
which indicates the unification~\cite{tmdtw3} of two mechanisms for SSAs in hard processes.
We point out that the relation in Eq.~(\ref{relation}) has also been verified in Ref.~\cite{Kang:2010hg}
in diquark model calculation.

\section{Interaction-independent T-odd twist-3 distributions}

As a comparison, we also calculate the {\it interaction-independent}
twist-3 T-odd TMD distributions in the same model.
As shown in Ref.~\cite{Bacchetta:2006tn}, the interaction-dependent and
interaction-independent twist-3 distributions satisfy model-independent
relations provided by the equation of motion (EOM) for the quark field.
It is interesting to check wether these relations (see Eqs.(3.64-3.72) in
Ref.~\cite{Bacchetta:2006tn}) for T-odd distributions hold in the
diquark model.

Up to order $1/Q$, the quark-quark correlator can be decomposed as
~\cite{Bacchetta:2006tn,Goeke:2005hb}:
\begin{align}
\Phi(x,p_T) &=\ldots + \frac{M}{2 P^+}\,\biggl\{e
- i\,{e_s} \,\gamma_5
- {e_{T}^\perp}\, \frac{\eps_T^{\rho \sigma} p_{T\rho}^{}\slim
  S_{T\sigma}^{}}{M}
\nonumber \displaybreak[0] \\ & \quad
+ f^\perp\, \frac{\pslash_T}{M}
- {f_T'}\,\epsilon_T^{\rho\sigma} \gamma_\rho^{}\slim S_{T \sigma}^{}
- {f_s^{\perp}}\,\frac{\eps_T^{\rho \sigma}
    \g_{\rho}^{}\slim p_{T \sigma}^{}}{M}
\nonumber \displaybreak[0] \\ & \quad
+ g_T'\, \gamma_5\Sslash_T
+ g_s^{\perp} \gamma_5 \frac{\pslash_T}{M}
- {g^\perp} \g_5\,\frac{\eps_T^{\rho \sigma}
    \g_{\rho}^{}\slim p_{T \sigma}^{}}{M}
\nonumber \displaybreak[0] \\[0.2em] & \quad
+ h_s\,\frac{[\nslash_+, \nslash_-]\gamma_5}{2}
+ h_T^{\perp}\,\frac{\bigl[\Sslash_T, \pslash_T \bigr]\gamma_5}{2 M}
+ i \, {h} \frac{ \bigl[\nslash_+, \nslash_- \bigr]}{2}
\biggr\} \,,
\label{eq:phi}
\end{align}
where we only write the terms containing twist-3 distributions, and we use
ellipse to denote the twist-two terms that have not been considered in this
letter. The leading-twist T-odd and T-even TMDs have also been calculated
in the diquark model, and a complete calculation can be found in
\cite{Bacchetta:2008af}.
 The distribution functions in (\ref{eq:phi}) can be deduced from the traces of the correlator
$\Phi(x, p_T)$:
\begin{align}
\Phi^{[1]} & =
 \frac{M}{P^+} \bigg[ e
%(x,{p}_{T}^{2})
 - \frac{\eps_{T}^{\rho\sigma} p_{T \rho}^{}\slim S_{T \sigma}^{}}{M} \,
   e_{T}^{\perp}
%(x,{p}_{T}^{2})
 \bigg] ,
\displaybreak[0] \\
\Phi^{[i \gamma_5]} & =
 \frac{M}{P^+} \bigg[ S_L \slim e_{L}
%(x,{p}_{T}^{2})
 - \frac{{p}_{T} \cdott {S}_{T}}{M} \,
   e_{T}
%(x,{p}_{T}^{2})
 \bigg] ,\displaybreak[0]
\\
\label{e:gi}
\Phi^{[\gamma^{\alpha}]} & =
 \frac{M}{P^+} \bigg[
 - \eps_{T}^{\alpha\rho} S_{T \rho}^{} \,
   f_{T}
%(x,{p}_{T}^{2})
 - S_L \,\frac{\eps_{T}^{\alpha\rho} p_{T \rho}^{}}{M}\,
   f_{L}^{\perp}
%(x,{p}_{T}^{2})
\nonumber \displaybreak[0] \\ & \qquad
 -  \frac{p_T^{\alpha}\slim p_T^{\rho}
        -\frac{1}{2}\,{p}_T^{2}\,g_{T}^{\alpha\rho}}{M^2}
        \,\eps^{}_{T \rho\sigma}\slim S_{T}^{\sigma}\,
   f_{T}^{\perp}
%(x,{p}_{T}^{2})
 + \frac{p_{T}^{\alpha}}{M} f^{\perp}
%(x,{p}_{T}^{2})
 \bigg] ,\displaybreak[0]
\\
\label{e:gi5}
\Phi^{[\gamma^{\alpha}\gamma_5]} & =
 \frac{M}{P^+} \bigg[
 S_{T}^{\alpha} \, g_{T}
%(x,{p}_{T}^{2})
 + S_L \, \frac{p_{T}^{\alpha}}{M}\, g_{L}^{\perp}
%(x,{p}_{T}^{2})
\nonumber \displaybreak[0] \\ & \qquad
 - \frac{p_{T}^{\alpha}\slim p_{T}^{\rho}
     -\frac{1}{2}\,{p}_T^{2}\,g_T^{\alpha\rho}}{M^2}
        \,S_{T \rho}^{}\,
   g_{T}^{\perp}
%(x,{p}_{T}^{2})
- \frac{\eps_{T}^{\alpha\rho} p_{T \rho}^{}}{M} \,
  g^{\perp}
%(x,{p}_{T}^{2})
 \bigg] ,\displaybreak[0]
\\
\Phi^{[i\sigma^{\alpha\beta}\gamma_5]} & =
 \frac{M}{P^+} \bigg[ \frac{S_{T}^{\alpha}\slim p_{T}^{\beta}
                     - p_{T}^{\alpha}\slim S_{T}^{\beta}}{M} \,
   h_{T}^{\perp}
%(x,{p}_{T}^{2})
 - \eps_{T}^{\alpha\beta} \, h
%(x,{p}_{T}^{2})
 \bigg] ,\displaybreak[0]
\\
\Phi^{[i\sigma^{+-}\gamma_5]} & =
 \frac{M}{P^+} \bigg[ S_L \slim h_{L}
%(x,{p}_{T}^{2})
 - \frac{{p}_{T} \cdott {S}_{T}}{M} \,
   h_{T}
%(x,{p}_{T}^{2})
 \bigg]\,,
\end{align}
Here  $\Phi^{[\Gamma]} = \half \Tr [\slim
\Phi\, \Gamma\slim]$.
The eight T-even twist-3 TMDs have been
calculated in the diquark model~\cite{Jakob:1997wg} and the Bag model~
\cite{Avakian:2010br}.
The T-odd twist-3 TMDs has been studied in Ref.~\cite{Gamberg:2006ru}, where
an explicit calculation for $g^\perp$ in the scalar diquark model showing
that the T-odd TMD distributions are divergent when one chooses a
point-like nucleon-quark-diquark coupling.
In order to compare to the interaction-dependent twist-3 distributions
listed in Eqs.~(\ref{etperp}) to (\ref{ft}), again we use a dipole form factor for the
nucleon-quark-diquark coupling to obtain finite results to order $\mathcal{O}(e_qe_s)$,
as also suggested in Ref.~\cite{Gamberg:2006ru}.

The expressions for the eight T-odd distributions in the scalar diquark model are
listed as follows
\begin{align}
 e_T^\perp(x,p_T^{\,2})
 &=-{A \over 2(1-x)} { (1-x)\Lambda^2+(1+x)M_s^2 -(1-x)(1-2x)M^2\over L (L +\vec p_T^{\,2})^3},\displaybreak[0] \\
 e_L (x,p_T^{\,2})&= {A\over 2(1-x)}{(xM+m)\over M }{L-\vec p_T^{\,2}\over L (L+\vec p_T^{\,2})^3},\displaybreak[0] \\
e_T(x,p_T^{\,2})
&=-{A \over 2(1-x)} {
2(1-x)Mm+ (1-x)M^2  -(1+x)M_s^2 -(1-x)\Lambda^2\over L (L +\vec p_T^{\,2})^3},\displaybreak[0] \\
h (x,p_T^{\,2})& =-{A\over 2(1-x)}{(xM+m)\over M }{L-\vec p_T^{\,2}\over L (L+\vec p_T^{\,2})^3},\displaybreak[0] \\
g^\perp (x,p_T^{\,2})&
 =-{A \over 2(1-x)} { (1-x)\Lambda^2+(1+x)M_s^2 -(1-x)(1-2x)M^2\over L (L +\vec p_T^{\,2})^3},\displaybreak[0] \\
f_L^\perp
  (x,p_T^{\,2})&=-{A\over 2(1-x)} {(1-x)((1+2x)M^2 +2mM -\Lambda^2)-(1+x)M_s^2\over L (L+\vec p_T^{\,2})^3},\displaybreak[0] \\
f_T^\perp(x,p_T^{\,2})&=0,\displaybreak[0] \\
f_T (x,p_T^{\,2})&= f_T^\prime(x,p_T^{\,2})=-{A\over 2(1-x)}{(xM+m)\over M }{L-\vec p_T^{\,2}\over L (L+\vec p_T^{\,2})^3}.
\end{align}
Among the above results, the calculation of $g^\perp$ is already given in
Ref.~\cite{Gamberg:2006ru}, while the others are new results.
In the scalar diquark model, we find that $f_T^\perp$ is zero.
We have checked that the inclusion of an (axial-)diquark yields
non-zero results for $f_T^\perp$.

Combining the results for the Sivers function and the Boer-Mulders function
in the same model~\cite{Bacchetta:2003rz}, we find that the
interaction-dependent and interaction-independent twist-3 TMDs
in the diquark model do not satisfy the EOM relations.
The reason for the disagreement between the diquark model calculations
and the EOM relations is that our calculation is model dependent, while EOM
relations in Ref.~\cite{Bacchetta:2006tn} is derived from the genuine QCD;
and we have only considered the lowest order contribution to the T-odd TMDs.

Although our calculation is model dependent, it is interesting to point out
that our results show that the following relations are satisfied:
\begin{eqnarray}
\int d^2 \vec p_T f_T(x,\vec p_T^{\, 2}) =0\,,~~\int d^2 \vec p_T e_L(x,\vec p_T^{\, 2}) =0\,,
~~\int d^2 \vec p_T h(x,\vec p_T^{\, 2}) =0\,. \label{integ}
\end{eqnarray}
This can be seen from the fact that the following integration yields zero result:
\begin{eqnarray}
\int d^2 \vec p_T {L-\vec p_T^{\, 2}\over L (L+\vec p_T^{\, 2})^3} =0\,.
\end{eqnarray}
The results showing in Eq.~(\ref{integ}) agree with the general constraint of
time-reversal invariance on the integrated quark correlator $\Phi(x)$, that
is, that the integrated twist-3 T-odd distributions should vanish~
\cite{Bacchetta:2006tn,Goeke:2005hb}.

\section{Summary}
In summary, we have studied the possibility to calculate the T-odd
interaction-dependent twist-3 quark distributions in the spectator
diquark model.
We find that in the lowest order approximation (neglecting the gauge-links
in the quark-gluon-quark correlator), we can obtain non-zero results for the
eight T-odd interaction-dependent twist-3 quark TMDs.
In the calculation we apply a dipole form factor for the nucleon-quark-diquark
coupling, instead of a point-like coupling, to avoid the light-cone divergences.
We also calculate the interaction-independent T-odd twist-3 TMDs in the same
model for comparison.
In the order $\mathcal{O}(e_qe_s)$ in the diquark model,
we verify the identity between the ETQS function $T_{q,F}(x,x)$ and
the first transverse moments of the Sivers function
$f_{1T}^{\perp(1)}(x)$.
However, we find that in the same model, the interaction-dependent
and interaction-independent T-odd twist-3 TMDs
do not satisfy the relations provided by the equation of motion
for the quark field.
The diquark calculation also shows that integrated T-odd distributions
$f_T(x)$, $e_L(x)$ and $h(x)$ vanish, as required by time
reversal-invariance for the integrated distributions.
Our study demonstrates the applicability and the limitations of the
diquark model in the understanding of the parton structure of the
nucleon at the twist-3 level.

{\bf Acknowledgements} This work is supported by NSFC (China) Project No.~11005018,
by FONDECYT (Chile) Project No.~1100715, by Project Basal FB0821, and by Teaching
and Research Foundation for Outstanding Young Faculty of Southeast University.

\end{document}